%% file: main.tex
\documentclass[sigconf]{acmart}

\input{macro.tex}
\begin{document}

\title{An Approach to Detecting Bugs in Pattern Based Bug Detectors}


 \author{Junjie Wang$^{1,2,3}$,  Yuchao Huang$^{1,3}$, Song Wang$^4$, Qing Wang$^{1,2,3,*}$}
 \affiliation{
  \position{$^1$Laboratory for Internet Software Technologies, $^2$State Key Laboratory of Computer Sciences, }
    \department{Institute of Software Chinese Academy of Sciences, Beijing, China; \\
   $^3$University of Chinese Academy of Sciences, Beijing, China; $^*$Corresponding author}
   \country{$^4$Lassonde School of Engineering, York University, Canada;} 
 }
 \email{{junjie,yuchao,wq}@iscas.ac.cn,  wangsong@eecs.yorku.ca}         


\begin{abstract}
\input{sec/abstract}

\end{abstract}


\maketitle

\input{sec/introduction}

\input{sec/methodology}

\input{sec/result}

\input{sec/discussion}
\input{sec/related}

\input{sec/conclusion}


\balance
\bibliographystyle{ACM-Reference-Format}
\bibliography{reference}

\end{document}

%% file: macro.tex
\usepackage{soul}
\usepackage{multirow}
\usepackage{epstopdf}
\usepackage{hyperref}
\usepackage{listings}
\usepackage{fancybox}
\usepackage{graphicx}
\usepackage{subcaption}
\usepackage{color}
\usepackage{paralist}
\usepackage{ragged2e}
\usepackage{enumitem}
\usepackage{algorithm}
\usepackage{algorithmic}
\usepackage[utf8]{inputenc}
\usepackage{listings}

\usepackage[framemethod=TikZ]{mdframed}
\usepackage{tcolorbox}
\makeatletter
\newcommand{\mybox}[1]{%
  \setbox0=\hbox{#1}%
  \setlength{\@tempdima}{\dimexpr\wd0+13pt}%
  \begin{tcolorbox}[boxrule=0.5pt, colback=white, arc=4pt,
      left=6pt,right=6pt,top=6pt,bottom=6pt,boxsep=0pt]
    #1
  \end{tcolorbox}
}
\usepackage{tikz}
\usepackage{pgfplots}
\usetikzlibrary{pgfplots.statistics,calc}
\usepackage{color}
\usepackage{xcolor}
\usepackage{array}
\usepackage{amsmath}
\usepackage{centernot}
\usepackage{xspace}
\usepackage{url}
\usepackage{verbatim}
\usepackage{wrapfig}
\usepackage{tabularx}
\definecolor{javared}{rgb}{0.6,0,0} 
\definecolor{javagreen}{rgb}{0.25,0.5,0.35} 
\definecolor{javapurple}{rgb}{0.5,0,0.35} 
\definecolor{javadocblue}{rgb}{0.25,0.35,0.75} 

\newcommand{\incon}{7,633}
\newcommand{\bug}{46}
\newcommand{\bugfp}{8}
\newcommand{\bugfn}{38}
\newcommand{\bugfni}{29}
\newcommand{\bugfnd}{9}
\newcommand{\cbug}{24}
\newcommand{\pattern}{13}

%% file: sec/abstract.tex
Static bug finders have been widely-adopted by developers to find bugs in real-world software projects. 
They leverage predefined heuristic static analysis rules to scan source code or binary code of a software project, and report violations to these rules as warnings to be verified. 
However, the advantages of static bug finders are overshadowed by such issues as uncovered obvious bugs, false positives, etc.
To improve these tools, many techniques have been proposed to filter out false positives reported or design new static analysis rules.
Nevertheless, the under-performance of bug finders can also be caused by the incorrectness of current rules contained in the static bug finders, which is not explored yet.
In this work, we propose a differential testing approach to detect bugs in the rules of four widely-used static bug finders, i.e., SonarQube, PMD, SpotBugs, and ErrorProne, and conduct a qualitative study about the bugs found. 
To retrieve paired rules across static bug finders for differential testing, we design a heuristic-based rule mapping method which combines the similarity in rules' description and the overlap in warning information reported by the tools. 
The experiment on 2,728 open source projects reveals {\bug} bugs in the static bug finders, among which {\cbug} are fixed or confirmed and the left are awaiting confirmation. 
We also summarize {\pattern} bug patterns in the static analysis rules based on their context and root causes, which can serve as the checklist for designing and implementing other rules and/or in other tools. 
This study indicates that the commonly-used static bug finders are not as reliable as they might have been envisaged. 
It not only demonstrates the effectiveness of our approach, but also highlights the need to continue improving the reliability of the static bug finders.

%% file: sec/introduction.tex
\section{Introduction}
\label{sec:intro}

The increasing complexity of modern software systems has complicated both the development of new software features and the maintenance of source code.
Techniques that can detect and reduce bugs are very beneficial to help developers improve software quality. 
To achieve this goal, static bug finders that analyze code characteristics without program execution such as Sonarqube~\cite{campbell2013sonarqube} and Findbugs~\cite{hovemeyer2004finding}, have been widely used to find bugs in software~\cite{bessey2010few,Ayewah2008using,Sadowski2018lessons,smith2019how}. 
These tools mainly leverage predefined heuristic analysis rules to scan source code or binary code of a software project, and report violations to these rules as warnings to be verified. Most of static bug finders can infer a wide variety of bugs, security vulnerabilities, and bad programming practices~\cite{flanagan2002extended,wang2018is,smith2019how}. 

Previous studies have shown that static bug finders can help in detecting software defects faster and cheaper than human inspection or software testing~\cite{beller2019developer,johnson2013why}.
They have been widespread adopted by professional software developers, and regularly integrated in contemporary open source projects and commercial software organizations~\cite{Kern2019integrating,Zampetti2017how,Panichella2015would}. 
For example, Errorprone and Infer has been automatically applied to code changes to support manual code review at Google and Facebook, respectively \cite{Edward2012building}. 
However, the advantages of static bug finders are overshadowed by such issues as uncovered obvious bugs \cite{Vassallo2020how,habib2018how}, false positives \cite{johnson2013why,lenarduzzi2020sonarqube}, etc.
To improve these tools and enhance their usability, different lines of studies have been proposed. 

Several researchers proposed to utilize prioritization strategies to make it easier for developers to spot the more actionable warnings~\cite{heckman2008establishing,kim2007which,liang2010automatic,heckman2011systematic,wang2018is}.
Other researchers employed feedback-based rule design for mitigating false positives~\cite{Sadowski2015Tricorder,Nam2019bug}, e.g., Errorprone used unactionable warnings labeled by developers \cite{Sadowski2015Tricorder}, and FeeFin refined the rules by the development practice of open source projects \cite{Nam2019bug}. 
Another line of studies focused on designing project-specific rules that mined from specific projects to improve the detection accuracy~\cite{li2005PRminer,Livshits2005dynamine,jin2012understanding,hanam2016discovering,Chen2017characterizing,bian2018NARminer}, e.g., PR-Miner~\cite{li2005PRminer} and NAR-Miner~\cite{bian2018NARminer} mined programming rules and detected violations with frequent itemset mining algorithms.  
All above mentioned practices focused on improving current static bug finders by adjusting the warning results based on existing static analysis rules or designing new rules.
However, the under-performance of bug finders can also be caused by the incorrectness of current rules contained in these static bug finders, which is not explored yet. 

In this work, to assess the correctness of static analysis rules of static bug finders, we propose a differential testing approach to detect bugs in the rules of four widely-used static bug finders, i.e.,  SonarQube, PMD, SpotBugs, and ErrorProne, and conduct a qualitative study about the bugs found in these tools. 

The assumption of our approach is that static analysis rules from different bug finders that target at the same type of bugs should have consistent bug detection results (i.e., warnings) on the same inputted software projects.  
Specifically, our work starts from retrieving \textit{paired rules} that target at the same type of bugs among different static bug finders.
We then run the static bug finders on a large set of experimental projects and check whether there are inconsistencies between the reported warnings of these paired rules.
To retrieve paired rules across static bug finders for differential testing, we design a heuristic-based rule mapping method.
which combines the similarity in rules' description and the overlap in warning information reported by the tools. 


To evaluate our approach, for the four examined bug finders, we treat SonarQube and PMD as a pair (which scan source code of software projects to detect bugs), while SpotBugs and ErrorProne as a another pair (which scan binary code of software projects to detect bugs).
Our heuristic-based rule mapping method retrieves 74 pairs of rules for SonarQube and PMD, while 30 pairs of rules are retrieved for SpotBugs and ErrorProne. 
We use 2,728 open source projects from an existing publicly available dataset as the experimental subjects for detecting the inconsistencies in the warnings.
Results show that {\incon} inconsistencies between the paired rules from different static bug finders have been revealed.

We then apply descriptive coding, a qualitative analysis method, on the detected inconsistencies to identify the buggy rules in the static bug finders and categorize these bugs to derive bug patterns. 
{\bug} bugs in the static analysis rules across the four static bug finders are found, among which {\bugfp} are bugs in the implementations of rules that cause them to generate false positives, {\bugfn} are bugs that cause them to miss detect true bugs (i.e., false negatives).
We further summarize {\pattern} bug patterns in the static analysis rules based on the bugs' context and root causes. 
For example, bug pattern \textit{fail in multiple calling operations} denotes the static analysis rules would fail to warn the suspicious code when involving the multiple calling operations as \textit{json.exception().printStackTrace()} (work for single calling operation as \textit{exp.printStackTrace()}).
These bug patterns can serve as the checklists for developers when designing and implementing other static analysis rules and/or in other static bug finders. 
We also localize these bugs and summarize three types of typical faults in the implementation of these static bug finders. 

To evaluate the usefulness of this study, we report these found bugs to the development team, and {\cbug} is fixed or confirmed and the left are awaiting confirmation.
This study indicates that the commonly-used static bug finders are not as reliable as they might have been envisaged. 
It 
highlights the need to continue improving the reliability of the static bug finders, and suggests the feasibility of utilizing differential testing on these static bug finders. 

This paper makes the following contributions:

\begin{itemize}
\item We conduct the first differential testing on four widely-used static bug finders, which is the first work on testing the correctness of static analysis rules in static bug finders to the best of our knowledge.

\item Our study finds {\bug} bugs about the implementation or design of static analysis rules, among which {\cbug} are fixed/confirmed\footnote{Details are listed in \url{https://github.com/wuchiuwong/Diff-Testing-01}.}.


\item We propose a heuristic-based static analysis rule mapping method to retrieve paired rules that target at the same types of bugs across different static bug finders.


\item We summarize {\pattern} bug patterns in the static analysis rules based on their context and root causes, which can serve as the checklist for designing and implementing other rules.

\end{itemize}


%% file: sec/methodology.tex
\section{Methodology}
\label{sec_method}

\subsection{Examined Static Bug Finders}
\label{subsec_method_tool}
In this study, we explore the correctness of static analysis rules for four popular open source static bug finders as listed below.

1) \textbf{\textit{SonarQube}} is one of the most widely adopted static bug finders that leverages pre-defined static analysis rules to help find bugs in the context of continuous integration. 
It supports more than 20 programming languages and has been adopted by more than 85,000 organizations or software projects.  
SonarQube provides developers with its own analysis rules and also incorporates rules from other popular bug finders, e.g., CheckStyle, PMD, and FindBugs. 
In this study, we only experiment with SonarQube's own rules, i.e., 545 Java related rules in its rule repository\footnote{\url{https://rules.sonarsource.com/java}}.  

2) \textbf{\textit{PMD}} is a source code analyzer maintained by open community. 
It finds common programming flaws like unused variables, empty catch blocks, and unnecessary object creation, etc. 
It supports Java, JavaScript, PLSQL, Apache Velocity, XML, etc. We include all its 304 Java related rules for experiment\footnote{\url{https://pmd.github.io/latest/pmd_rules_java.html}}.

3) \textbf{\textit{SpotBugs}} is the spiritual successor of the pioneering FindBugs tool \cite{Hovemeyer2007finding}, carrying on from the point where it left off with support of its community. 
It is a bug finder which uses static analysis to look for bugs in Java code.  
It was originally developed by the University of Maryland, and has been downloaded more than a million times.
We use all the 449 Java related rules for experiment\footnote{\url{https://spotbugs.readthedocs.io/en/latest/bugDescriptions.html}}. 
 
4) \textbf{\textit{ErrorProne}} is a static bug finder for Java that catches common programming mistakes at compile-time.
It is developed by Google and is integrated into their static analysis ecosystem \cite{Sadowski2018lessons}.
We experiment with all its 333 Java related rules\footnote{\url{https://errorprone.info/bugpatterns}}.

The first two static bug finders work on the source code of software projects, while the last two tools require the compiled binary code of software projects.  
Since this difference might lead to the variations in the marked line of the suspicious code, we group the first two tools (i.e., SonarQube and PMD) as a pair and the last two tools (SpotBugs and ErrorProne) as the second pair to conduct the differential testing. 
Note that, this study focuses on the static bug detection of Java projects with Java related static analysis rules, which is one of the most commonly-used programming languages. 

\subsection{Experimental Projects}
\label{subsec_method_projects}
In order to fully explore the static analysis rules, we need a large set of projects whose source code is available and compilable (for collecting binary code).
We also expect these projects having flaws to cover as many static analysis rules as possible, so that the static bug finders can be triggered and the inconsistencies in their warning results can be potentially revealed. 

To satisfy all above requirements, we choose to use the 50K-C projects repository \cite{Martins201850K}, which contains 50,000 Java projects crawled from GitHub.
Each project is attached with the dependencies required to compile it, and the scripts with which the projects can be compiled.
All of them are active open source projects, so that there should be dozens of flaws which can trigger the static analysis rules.

\input{figure/projects}

We randomly download 3,000 projects from the repository, then employ their provided building framework \textit{SourcererJBF} to compile the source code\cite{Martins201850K}.
272 projects could not be successfully compiled because of such errors as incompatible character set.
We use the remaining 2,728 projects with  1565 KLOC (Kilometer Lines Of Code) in total, for the following experiment. 
Figure \ref{fig:projects} presents the details of these projects with number of files in each project, average lines of source code per file in each project, and number of dependent jars for compiling the project.  
Note that, the maximum value of the first two series of data are respectively 1877 and 3370, and we cut off the figures to facilitate visualization.
There are an average of 32 files in an experimental project, and each file has an average of 115 lines of source code.

\subsection{Differential Testing of Static Analysis Rules}
\label{subsec_method_procedure}

The primary idea of this study is to detect bugs in static bug finders through differential testing, i.e., by providing the same input to different implementations of the same functionality and observing the inconsistencies between the implementations.
To achieve this goal, we treat each rule implemented in the static bug finder as a functionality of the bug finder, and treat the paired rules from different bug finders which target at detecting the same types of bugs as different implementations of the same functionality. 


As demonstrated in Figure \ref{fig:approach}, this study first retrieves the paired rules between two static bug finders (Section \ref{subsec_method_procedure_map}), i.e., two rules target at the same types of bugs, then detects the inconsistencies between the bug finders when respectively running the paired rules with the same inputted software projects (Section \ref{subsec_method_procedure_detect}).
Based on the inconsistencies, it identifies the buggy rules in the static bug finders and categorizes them (Section \ref{subsec_method_procedure_categorize}), meanwhile it also localizes the found bugs in the static bug finders (Section \ref{subsec_method_procedure_manual}).

\input{figure/approach}

\subsubsection{Retrieving Paired Rules}
\label{subsec_method_procedure_map}

As showed in Section~\ref{subsec_method_tool}, each examined static bug finder has hundreds of rules, and there would be tens of thousands candidate pairs, e.g., 165,680 (545 $\times$ 304) pairs for matching each rule of Snoarqube to each rule of PMD. 
Manually mapping such large number of rules from different bug finders could be time- and effort-consuming. 
Besides, rules from different tools are described differently, e.g., rule \textit{throwable.printStackTrace should not be called} in Sonarqube is described with 80 words with two code examples, while its paired rule \textit{AvoidPrintStackTrace} in PMD has only 7 terms with one code example, which further increases the difficulty of rule mapping.

To cope with the above circumstance,  we propose a heuristic-based rule mapping method which combines the similarity in rules' description and the overlap in warning information reported by the tools. 
In detail, we first choose the potential rule pairs with the similarity of rules based on their textual descriptions and accompanied code examples.  
For each candidate rule pair, we then check the detailed warning information reported by the rules and filter out the less possible rule pairs, in which two rules with larger degree of overlap have higher possibility to be a pair.
This is conducted with four \textit{mapping-rules} which will be described below.




\vspace{0.05cm}
\textbf{1) Choosing Potential Rule Pairs Based on Description Similarity}
\vspace{0.05cm}

Generally speaking, each static analysis rule in these static bug finders is described with three fields: \textit{title}, \textit{detailed description}, and \textit{code examples}, in which title and description demonstrate what types of bugs the rule targets at and how the analysis works, while code examples present the positive and negative examples to show when the rule is triggered.
We concatenate the title and detailed description fields, and treat it as the textual content of the rule. 
To model the similarity between the descriptions of two rules from different aspects, we use three types of similarity metrics, i.e., \textit{term similarity} to measure the text similarity of the rule's textual content,  \textit{semantic similarity} to measure the semantic similarity of the rule's textual content, and \textit{code similarity} to measure the similarity of code examples. Details are as follows.

\textbf{\textit{Term similarity ($Term_{sim}$)}}.
The term similarity is measured with the Term Frequency and Inverse Document Frequency (TF-IDF).
It is one of the most popular features for representing textual documents in information retrieval. 
The main idea of TF-IDF is that if a term appears many times in one rule and a few times in the other rules, the term has a good capability to differentiate the rules, and thus the term has high TF-IDF value. 
Specifically, given a term $t$ and a rule $r$, $\mathit{TF(t,r)}$ is the number of times that term $t$ occurs in rule $r$, while $\mathit{IDF(t)}$ is obtained by dividing the total number of rules by the number of rules containing term $t$.
TF-IDF is computed as: $\mathit{TF-IDF(t,r)} = \mathit{TF(t,r)} \times \mathit{IDF(t)}$. 

With the above formula, the textual content of a rule $r$ can be represented as a TF-IDF vector, i.e., $r = (w_1, w_2, ..., w_n)$, where $w_i$ denotes the TF-IDF value of the $i^{th}$ term in rule $r$.
Then term similarity is calculated as the cosine similarity between the vectors of two rules.

\textbf{\textit{Semantic similarity ($Semt_{sim}$)}}.
The above mentioned TF-IDF similarity focuses on the similarity of rules considering the term matching.
We also employ word embedding feature, which concerns more on the relationship of terms considering the context they appear, to better model the semantic similarity of two rules. 
Word embedding is a popular feature learning technique in natural language processing where individual words are no longer treated as unique symbols, but represented as $d$-dimensional vector of real numbers that capture their contextual semantic meanings \cite{mikolov2013distributed,Bengio2003neural}.

We use the publicly available software\footnote{https://code.google.com/archive/p/word2vec/} to obtain the word embedding of a rule.
With the trained word embedding model, each word can be transformed into a $d$-dimensional vector where $d$ is set to 100 as suggested in previous studies \cite{yang2016combining,xu2016predicting,Wang2019images}.
Meanwhile a rule can be transformed into a matrix in which each row represents a term in the rule.
We then transform the rule matrix into a vector by averaging all the word vectors the rule contains as previous work did \cite{yang2016combining,xu2016predicting,Wang2019images}.
Specifically, given a rule matrix that has $n$ rows in total, we denote the $i^{th}$ row of the matrix as $r_{i}$ and the transformed rule vector $v_{d}$ is generated as follows:

\vspace{-0.1in}
\begin{equation}
\label{equation_3}
v_{d} = \frac{\sum_{i}r_{i}}{n}
\end{equation}

With the above formula, each rule can be represented as a word embedding vector, and semantic similarity is calculated as the cosine similarity between the vectors of two rules.

For training the word embedding model, we collect 100,000 java related questions and answers from StackOverflow.
We then combine these text with the rule description of the four examined tools, and utilize them for model training. 
The reason why we use these data is that previous studies have revealed that to train an effective word embedding model, a domain-specific dataset with large size is preferred.
The size of our training dataset is 101 Megabytes.

\textbf{\textit{Code similarity ($Code_{sim}$)}}.
The code examples of each rule contain the name of class and method targeted by the rule, which are indispensable sources of information for determining whether two rules share the same functionality.
We first extract the class name and method name of each rule, separate them into terms with camel-back notation following existing study \cite{link_camel_case}. 
We then compare the two set of terms of $rule_x$ and $rule_y$ to derive the code similarity with the following equation.

\vspace{-0.1in}
\begin{equation}
\label{equation_codesim}
Code_{sim} = \frac{\mathit{terms} \ in \ \mathit{rule_x} \cap \mathit{terms} \ in \ \mathit{rule_y}}{\mathit{terms} \ in \ \mathit{rule_x} \cup \mathit{terms} \ in \ \mathit{rule_y}}
\end{equation}

The final description similarity between two rules is calculated as follows.

\vspace{-0.1in}
\begin{equation}
\small
\label{equation_totalsim}
Description_{sim}=(\mathit{Term_{sim}}+\mathit{Semt_{sim}})\times \frac{ \mathit{Code_{sim}}+1}{2}
\end{equation}

We simply add the term similarity and semantic similarity together because existing researches suggested both of them are important \cite{Wang2019images,yang2016combining}.
The code similarity, which is smoothed considering it might be 0, can be seen as a filter by which if two rules share large portion of class and method names, they are more likely to be paired, otherwise they are less likely to be paired even they are similar in textual content. 

Based on the description similarity, we design \textit{mapping-rule a} to choose the potential static analysis rule pairs.

\begin{itemize}
\item \textbf{\textit{(Mapping-rule a) Retrieving pairs of mutual top-N similarity.}}
If $rule_b$ is within $rule_a$'s \textit{top-N} most similar rules and vice versa, we treat $rule_a$ and $rule_b$ as a candidate pair for further investigation, where $rule_a$ and $rule_b$ are respectively from a pair of static bug finders. 
\end{itemize}

\vspace{0.05cm}
\textbf{2) Filtering Out Less Possible Rule Pairs Based on Warning Information} 
\vspace{0.05cm}

We then run each static bug finder on the experimental projects collected in Section~\ref{subsec_method_projects}. 
For each reported warning, we record the triggered static analysis rule, the warned line(s) of code, the method and file where the warning occurs.

Based on the reported warning information, we then design \textit{mapping-rules (b,c,d)} to jointly filter out the less possible rule pairs from the potential rule pairs generated with \textit{mappping-rule (a)}.
These three mapping-rules are designed considering the following two assumptions: 1) the paired rules should have large degree of overlap in their warnings; 2) considering one of the paired rules might have bugs, we allow their reported warnings can be partially overlapped. 

\begin{itemize}
\item \textbf{\textit{(Mapping-rule b) Pruning with one-to-one pair.}} 
If the percentage of overlaps of the warned lines reported by $rule_a$ and $rule_b$ exceeds 80\% of warned lines from each of the rules, we assume the candidate pair has extremely high possibility being the paired rules. 
Note that, to simplify our approach, we focus on one-to-one mapping among two sets of rules from different bug finders, thus we remove other candidate pairs related with $rule_a$ and $rule_b$.
Another note is that, we choose the pair with the largest overlap ratio when multiple rule pairs satisfy the threshold. 

\item \textbf{\textit{(Mapping-rule c) Pruning with difference in warning trigger times.}}
If the difference of warning trigger times for $rule_a$ and $rule_b$ exceeds 20 times, we assume these two rules can hardly related with the same functionality and remove the candidate pair. 

\item \textbf{\textit{(Mapping-rule d) Pruning with difference in warning file.}}
If the overlap of warned files by $rule_a$ and $rule_b$ is lower than 2\%, we assume these two rules can hardly related with the same functionality and remove the candidate pair. 

\end{itemize}


We find a large portion of rules from SpotBugs and ErrorProne do not have code examples and exert a lower similarity; thus for this tool pair, we set \textit{N} as 5 in \textit{mapping-rule a}, while set it as 3 for another tool pair (i.e., SonarQube and PMD). 
Other parameters in the mapping rules is determined empirically, which aims at automatically retrieving a reasonable number of candidate pairs for manual inspection.
We will mention in the threats to validity that there do exist one-to-many mappings, however for facilitating the proposed differential testing, we only focus on one-to-one mapping between rules from different tools.  

\vspace{0.05cm}
\textbf{3) Retrieving Final Rule Pairs Manually} 
\vspace{0.05cm}

For all the remaining candidate paired rules after the above four mapping-rules, we conduct a manual check to finally determine the paired rules.
In detail, the first three authors independently check the candidate pairs, and determine whether the two rules have the same functionality based on the rules' description and code examples. 
The results of their independent mapping have a Cohens kappa of 0.87, which is a substantial level of agreement \cite{mchugh2012interrater}. 
They then discuss the disagreement online until the final consensus is reached.

\subsubsection{\textbf{Detecting Inconsistencies}}
\label{subsec_method_procedure_detect}

Based on the retrieved paired rules, we check the inconsistencies in the warnings generated by the paired rules when running the static bug finders, i.e., whether the warnings reported by paired rules mark the same place in the source code.
Since different rules would highlight the warnings at different granularities, e.g., a specific line of code or the whole method, we employ different criteria for the inconsistency detection for different paired rules. 

\textbf{\textit{Criterion 1}}, when both of the paired rules warn \textit{a specific line of code}, we check whether the warned file and warned line coincide with each other between the paired rules, and treat the case in which the warned lines are different as the inconsistency.

\textbf{\textit{Criterion 2}}, when one or two of the paired rules warn(s) the \textit{whole method}, e.g., rule \textit{ReturnEmptyArrayRatherThanNull} of PMD marks the entire method while its paired rule (\textit{Empty arrays and collections should be returned instead of null}) in SonarQube only marks the \textit{return} line (as the example shown below), we check whether the warned methods coincide with each other between the paired rules, and treat the case in which the warned methods are different as the inconsistency.

\begin{lstlisting}
private float[] getReactionFlyShip(Ship ship) { //warn by PMD
    float box[] = new float[8];
    box[0] = ship.getAngle() / (2f * 3.14159f);
    ...
    return null;   //warn by SonarQube
}
\end{lstlisting}


\subsubsection{\textbf{Identifying and Categorizing Bugs}}
\label{subsec_method_procedure_categorize}

We apply a qualitative analysis method called descriptive coding \cite{descriptivecoding}\cite{rahman2019seven} on the detected inconsistencies to identify the buggy rules in the static bug finders and categorize these bugs to derive bug patterns. 
Figure \ref{fig:approach_bug_pattern} demonstrates an example of how we analyze the inconsistencies and summarize bug patterns.

The detected inconsistencies are delivered to the first three authors respectively, and each of them manually checks them and identifies the bugs.
Specifically, they first determine which of the paired rules is buggy, and whether the inconsistency involves a false negative bug (i.e., suspicious code is not warned) or a false positive bug (i.e., normal code is wrongly warned).
We also find some inconsistencies are caused by the the imperfectness in the rule definition (see Section \ref{subsec_result_bugs_FN_def}), therefore the authors also check the description of these static analysis rules to determine whether the bug is caused by the inaccurate \textit{implementation}, or mainly because of the imperfectness in rule \textit{definition}, to provide a more comprehensive view of these buggy rules.
The authors then examine the context and root causes of the bug, and summarize bug patterns of these buggy rules, as shown in Table \ref{tab:bug_rule_implement}. 
The disagreement of the above analysis is discussed until common consensus is reached. 

The categorization of the bug patterns might subject to the author's personal judgement. 
We mitigate this limitation by recruiting two independent workers who are not authors of the paper. 
These two workers, with more than five years background in software development, independently evaluate if the derived bug pattern is meaningful and correct.
We provide them with the bug pattern name, the involved buggy rule, random-chosen one consistent warning and two inconsistent warnings for each buggy rule.
The two workers independently determined if each of the buggy rule belongs to the bug pattern, and whether the bug pattern is meaningful. 
Both workers agree with the categorization and most of the bug patterns. 
The only disagreement is the name of \textit{P6. Fail in unnecessary brackets} whose original name is \textit{Fail in AST change by unnecessary brackets}.
This high degree of agreement implies the quality of the derived bug patterns.


\input{figure/approach_pattern}

\subsubsection{\textbf{Localizing Bugs}}
\label{subsec_method_procedure_manual}

For the detected buggy static analysis rules, we further examine the source code of the static bug finders to localize the bugs. 
For PMD, some rules are implemented with XPath technique~\cite{link_xpath_pmd}, while the others are implemented in JAVA, both of which are navigated in the XML configuration files. 
We use the class names to localize the rules' implementation, and examine the faults in the source code of static bug finders.
For other three static bug finders, we search the related JAVA file with corresponding rule name or rule id to localize the bugs. 

%% file: figure/projects.tex
\begin{figure}[t!]
  \centering
  \begin{subfigure}{0.16\textwidth}
    \includegraphics[width=2cm]{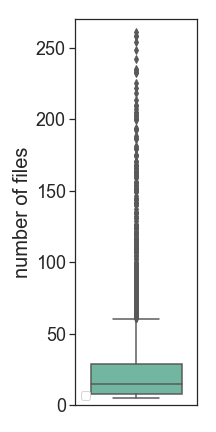}
	 \caption{\# files}	 
	 \label{fig:file}
  \end{subfigure}
  \hspace{-0.1in}
    \begin{subfigure}{0.16\textwidth}
    \includegraphics[width=2cm]{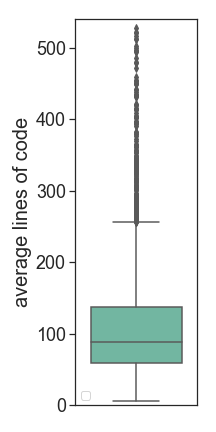}
	 \caption{\# LOC per file}	 
	 \label{fig:avelines}
  \end{subfigure}
  \hspace{-0.1in}
  \begin{subfigure}{0.16\textwidth}
    \includegraphics[width=2cm]{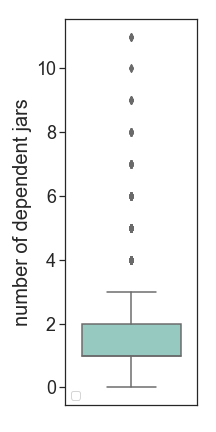}
	\caption{\# dependent jars}
   \label{fig:jars}
  \end{subfigure}  
  \caption{Details of experimental projects}
  \label{fig:projects}
\end{figure}

%% file: figure/approach.tex
\begin{figure*}[t!]
\centering
\includegraphics[width=13cm]{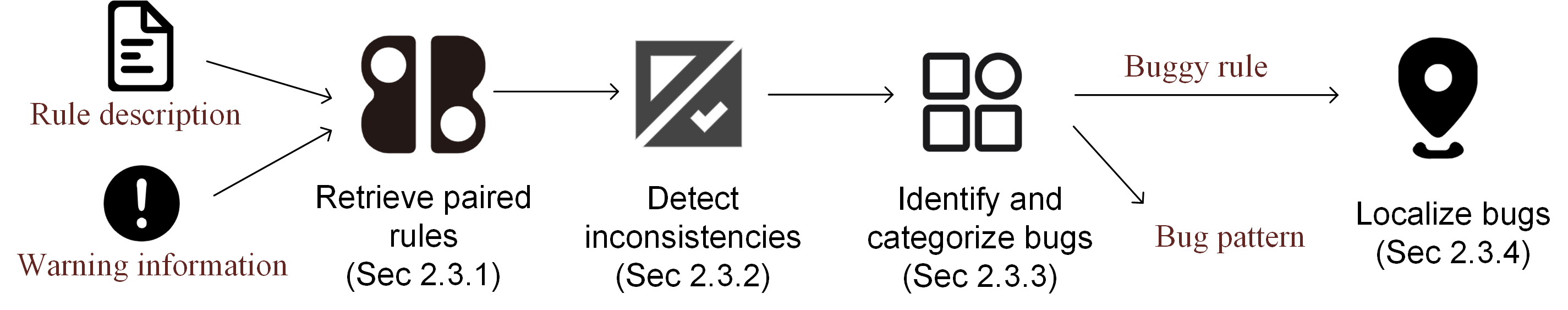}
\vspace{-0.05in}
\caption{Overview of differential testing of static analysis rules}
\label{fig:approach}
\vspace{-0.1in}
\end{figure*}

%% file: figure/approach_pattern.tex
\begin{figure*}[h!]
\centering
\includegraphics[width=17.5cm]{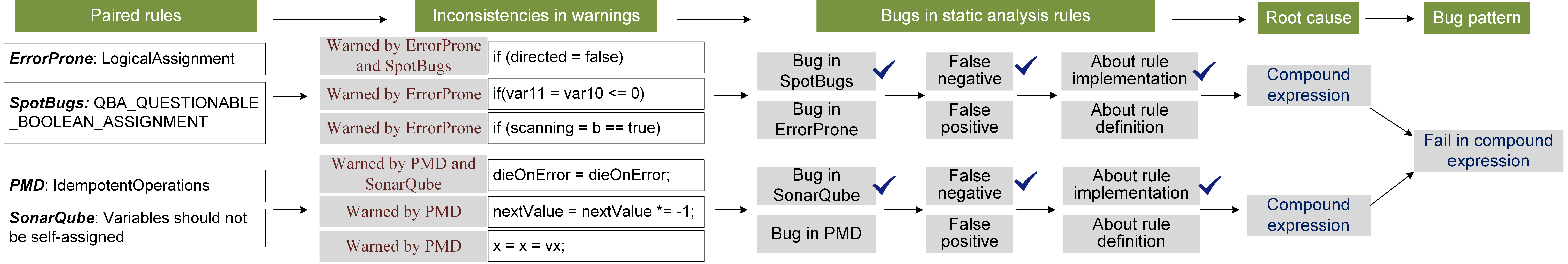}
\vspace{-0.05in}
\caption{An example of how we analyze the inconsistencies and summarize bug patterns}
\label{fig:approach_bug_pattern}
\vspace{-0.1in}
\end{figure*}

%% file: sec/result.tex
\section{Experimental Results}
\label{sec_results}

\subsection{Results of Rule Mapping}
\label{subsec_results_mapping}

We first present the results of how many rules of each bug finder are triggered and the warning trigger times after running the bug finders on our experimental projects, with results in Table \ref{tab_warning_info}.
We can see that 74\% to 89\% rules are triggered for SonarQube, PMD and SpotBugs, while for ErrorProne, 45\% of its rules are triggered.
We analyze the un-triggered rules of ErrorProne and find that most of them are related to the framework or package developed by Google itself, e.g., AutoValue, Dagger, etc., which are seldom used by general software projects, thus were not triggered.
On average, 73\% rules are triggered and each rule is triggered a median of 163 times, which suggests the generalizability of our experiments.  

\begin{table*}[ht!]
\caption{Statistics of warning information on experimental projects}
\centering
\footnotesize
\begin{tabular}{p{2cm}|p{2.2cm}|p{1.2cm}|p{1.2cm}|p{1.2cm}|p{1.2cm}|p{1.2cm}}
\hline
\multirow{2}{*}{\textbf{Tool}} & \multirow{2}{*}{\textbf{\% triggered rules}} & \multicolumn{5}{c}{\textbf{Warning trigger times per triggered rule}} \\ 
\cline{3-7}
& & \textbf{min} & \textbf{1-quarter} & \textbf{median} & \textbf{3-quarter} & \textbf{max} \\
\hline
\textbf{SonarQube} & 80\%  & 1 &  38  & 378  & 2,320 & 1,036,425 \\ 
\hline
\textbf{PMD} & 89\%  & 2 &  183  & 1,091  & 6,967 & 812,325 \\ 
\hline
\textbf{SpotBugs} & 74\%  & 1 &  7  & 45  & 216 & 9,049 \\ 
\hline
\textbf{ErrorProne} & 45\% & 1 &  7  & 36  & 137 & 77,896 \\ 
\hline
\hline
\textbf{Overall} & 73\% & 1 &  22  & 163  & 1,363 & 1,036,425 \\ 
\hline
\end{tabular}
\label{tab_warning_info}
\end{table*}

Following the rule mapping method described in Section \ref{subsec_method_procedure_map}, we retrieve the paired rules and list the results in Table \ref{tab_paired_rules}, in which we present the number of remaining rule pairs after applying \textit{mapping-rule (a-d)}.
74 rule pairs from SonarQube and PMD, and 30 rule pairs from SpotBugs and ErrorProne are finally determined as having identical functionality\footnote{These paired rules are listed in https://github.com/wuchiuwong/Diff-Testing-01.}. 
The hitting rate of our mapping method is 51.5\% (74/145 in Table \ref{tab_paired_rules}) for paired tools SonarQube and PMD, 45.4\% (30/66 in Table \ref{tab_paired_rules}) for paired tools SpotBugs and ErrorProne. 
This indicates we can find one mapped rule pairs by examining an estimate of two candidate pairs,  which is quite cost-effective considering the tremendous number of total candidate pairs. 

It is almost impossible to measure the recall of our retrieved rule pairs, considering the large number of candidate pairs. 
We construct a small-scale ground truth set to roughly evaluate the recall. 
In detail, we choose the rule pairs which 1) the overlap of warned files by two rules is 100\%; and 
2) the warning trigger times are larger than 10 to reduce the noise. 
Based on the filtered rule pairs, we conduct the manual check as previous section and obtain the paired rules. 
Since these rule pairs are determined solely based on the warning information, they can be treated as orthogonal with the rule pairs retrieved by the proposed mapping method which are firstly determined with description similarity. 
Based on the constructed ground truth rule pairs, 56\% (42/74) rule pairs for SonarQube and PMD are recalled, and 73\% (22/30) rule pairs for SpotBugs and ErrorProne are recalled. 
This further indicates the effectiveness of our proposed mapping method which can help find sufficient number of rule pairs with little human effort. 


\begin{table*}[ht!]
\caption{Results of rule mapping}
\centering
\footnotesize
\begin{tabular}{p{3.2cm}|p{2cm}|p{2cm}|p{2cm}|p{2cm}|p{2cm}|p{1.2cm}}
\hline
\textbf{Tool} & \textbf{Total candidate pairs} &  \textbf{Pairs after} \protect\newline \textbf{applying rule \textit{a}} & \textbf{Pairs after} \protect\newline \textbf{applying rule \textit{b}}  & \textbf{Pairs after} \protect\newline \textbf{applying rule \textit{c}} & \textbf{Pairs after} \protect\newline \textbf{applying rule \textit{d}} & \textbf{Final pairs} \\ \hline
\textbf{Sonarqube \& PMD}  & 165,680 & 424 & 367 & 252 & 145 & 74 \\ 
\hline
\textbf{ErrorProne \& SpotBugs}   & 149,517 & 432 & 387 & 264 & 66 & 30 \\ 
\hline
\end{tabular}
\label{tab_paired_rules}
\end{table*}

\subsection{Detected Bugs in Static Bug Finders}
\label{subsec_results_bugs}

Overall, {\incon} inconsistencies (i.e., sum of \textit{occurrence times} in Table \ref{tab:bug_rule_implement}, \ref{tab:bug_rule_define} and \ref{tab:bug_false_positive}) are revealed from the two pairs of static bug finders on our experimental projects.

Following the qualitative analysis method presented in Section \ref{subsec_method_procedure_categorize}, we examine these inconsistencies and identify {\bug} bugs in the rules of the four static bug finders, i.e., 10 in Sonarqube, 25 in PMD, 6 in SpotBugs, and 5 in ErrorProne, as shown in Table \ref{tab_detected_bugs}.
Among the bugs, {\bugfn} are false negative bugs (i.e., the suspicious code is not warned by the tool), while {\bugfp} are false positive bugs (i.e., the clean code is warned by the tool).  
In addition, for false negative bugs, {\bugfni} bugs are caused by rule implementation, while {\bugfnd} bugs are caused by rule definition; for false positive bugs, all of them are because of the rule implementation.
We discuss the three types of bugs in the following three subsections respectively. 

\begin{table*}[ht!]
\caption{Detected bugs on four static bug finders}
\centering
\footnotesize
\begin{tabular}{p{2cm}|p{4cm}|p{3cm}|p{2cm}|p{1.5cm}}
\hline
\textbf{Tool} & \textbf{False negative} \protect\newline\textbf{(about rule implementation)} & \textbf{False negative} \protect\newline \textbf{(about rule definition)} & \textbf{False Positive} & \textbf{Overall} \\ \hline
\textbf{SonarQube} & 6  & 3 &  1 & 10 \\ 
\hline
\textbf{PMD}  & 15 &  4 & 6 & 25 \\ 
\hline
\textbf{SpotBugs}  & 4 & 1 & 1 & 6 \\ 
\hline
\textbf{ErrorProne}  & 4 & 1 & 0 & 5 \\ 
\hline \hline
\textbf{Overall}  & 29 & 9 & 8 & 46 \\ 
\hline
\end{tabular}
\label{tab_detected_bugs}
\end{table*}

\input{tab/bug_rule_implement}

\subsubsection{\textbf{False Negative Bugs (about Rule Implementation)}}
\label{subsec_result_bugs_FN_imp}

For the {\bugfni} false negative bugs about rule implementation, seven bug patterns are summarized following the procedure described in Section \ref{subsec_method_procedure_categorize}.
Table \ref{tab:bug_rule_implement} demonstrates a summarized view of these bug patterns, with the illustrative example, the involved static analysis rules and the occurrence times of the bug (i.e., the number of inconsistencies for triggering the bug).
There are 3 remaining bugs, each of which belongs to a specific type, thus we put them in \textit{Others} category, and leave them for future exploration. 
We then present the detailed analysis of an example pattern to facilitate understanding.

\textbf{\textit{P2) Fail in compound expression.}}
The involved rules in this pattern cannot work with compound expressions (i.e., two or more operands in an expression). 
For example, rule \textit{QBA\_QUESTIONABLE\_} \textit{BOOLEAN\_ASSIGNMENT} from Spotbugs checks \textit{a literal boolean value (true or false) assigned to a boolean variable inside an if or while expression. Most probably this was supposed to be a boolean comparison using ==, not an assignment using =.}
Code example (a) shows the consistent case where both SpotBugs and ErrorProne can detect the suspicious code with the paired rules, while code example (b) shows the inconsistent case where only ErrorProne marks the suspicious code, i.e., the corresponding rule in SpotBugs has a bug.
The compound expression in code example (b) triggers the bug of this rule in SpotBugs. 

\noindent\textcolor{gray}{\footnotesize{(Code example a.) Warning reported by both SpotBugs and ErrorProne.}}
\begin{lstlisting}[]
//birker-fsm/fsm-master/src/fsm/EdgeFsm.java
public void setDirected(boolean directed) {
    if (directed = false) throw new IllegalArgumentException(""Fsm are always directed!""); //warn by SpotBugs and ErrorProne
}
\end{lstlisting}
\textcolor{gray}{\footnotesize{(Code example b.) Warning reported by ErrorProne only.}}
\begin{lstlisting}[]
//lunchza-VisualHDD/VisualHDD-master/Visual HDD/src/visual/gui/ProgramWindow.java
public void setScanStatus(boolean b) {
    if (scanning = b == true){	//mark only by ErrorProne
        scanning = true;}
    else if(scanning = b == false && canceled == true){  //warn only by Errorprone
        scanning = false;
	...
}
\end{lstlisting}

\input{tab/bug_rule_definition}

\subsubsection{\textbf{False Negative Bugs (about Rule Definition)} }
\label{subsec_result_bugs_FN_def}

For the {\bugfnd} false negative bugs about rule definition, three bug patterns (as shown in Table \ref{tab:bug_rule_define}) are summarized following the procedure described in Section \ref{subsec_method_procedure_categorize}.
The bugs in the above subsection are caused by the inaccurate \textit{implementation} of rules, while the bugs in this subsection are mainly because of the imperfectness in rule \textit{definition}. 
For example, we find the definition of some rules miss specific data types.
We separate them to remind the tool developers about the flaw in the design of these static analysis rules. 
We then present the detailed analysis of an example pattern to facilitate understanding.

\textbf{\textit{P8) Miss comparable method.}}
The rules in this pattern miss certain comparable method.
Take the rule \textit{UseLocaleWithCaseConversions} from PMD as an example. 
We present its rule description, as well as the description of its paired rule \textit{Locale should be used in String operations} of SonarQube as follows.

\begin{itemize}
\item \textit{PMD ({UseLocaleWithCaseConversions})}: When doing String:: \textit{toLowerCase()/toUpperCase()} conversions, use an explicit locale argument to specify the case transformation rules. 

\item \textit{SonarQube ({Locale should be used in String operations)}}: Failure to specify a locale when calling the methods \textit{toLowerCase(), toUpperCase() or format()} on String objects means the system default encoding will be used, possibly creating problems with international characters or number representations.
\end{itemize}

We can see that in the rule definition of PMD, only two string related methods are mentioned, while the third method \textit{format()} is included in the rule definition of SonarQube. 
The results from the inconsistent detection by running these two static bug finders on experimental projects confirm that \textit{String.format()} cannot be warned by PMD, which suggest the design of the rule \textit{UseLocaleWithCaseConversions} from PMD is incomplete and could be buggy.

\input{tab/bug_false_positive}

\subsubsection{\textbf{False Positive Bugs}}
\label{subsec_result_bugs_FP}

For the {\bugfp} false positive bugs, three bug patterns (as shown in Table \ref{tab:bug_false_positive}) are summarized following the procedure described in Section \ref{subsec_method_procedure_categorize}.
We then present the detailed analysis of an example pattern to facilitate understanding.

\textbf{\textit{P11) Poor handling of method with same name.}}
Bugs related with the rules in this pattern occur because they warn the correct method which shares the same method name (yet different method signatures) with the defined suspicious method. 
Take the rule \textit{Thread.notify()} from PMD as an example. This rule states \textit{its usually safer to call notifyAll() rather than notify() because the later one awakens an arbitrary thread monitoring the object when more than one thread is monitoring.} 
The following code examples first present the consistent case where both PMD and SonarQube can detect the suspicious code in example (a), followed by the inconsistent case where PMD wrongly highlights the normal code in example (b), i.e., bug in PMD.
We can see that although the method name is \textit{notify}, it is not \textit{Object.notify()} as defined in the rule, since these two methods have different method signatures.
PMD does not filter this special yet misleading case, which suggests a potential bug in the rule of PMD.

\noindent\textcolor{gray}{\footnotesize{(Code example a.) Warning reported by both SonarQube and PMD.}}
\begin{lstlisting}
//belaban-JGroups/JGroups-master/tests/other/org/jgroups/tests/TestToaOrder.java
public void memberFinished(Address addr) {
    synchronized (members) {
        members.remove(addr);
        if (members.isEmpty()) {
            members.notify();  //warn by PMD and SonarQube
        }
    }
}
\end{lstlisting}
\textcolor{gray}{\footnotesize{(Code example b.) Warning reported by PMD only.}}
\begin{lstlisting}
//mypsycho-SwingAppFramework/SwingAppFramework-master/src/main/java/org/mypsycho/beans/Injection.java	
private void injectSimple(Object bean, InjectionContext context){
    ...
    else if (toSet && !child.definition.isEmpty()) {
        getInjector().notify(getCanonicalName(),  "'" + child.definition + "' has been converted as null", null);  //warn only by PMD
    }
}
\end{lstlisting}

\subsection{Typical Faults in Static Bug Finders Causing Buggy Rules}
\label{subsec_result_faults}

Besides detecting the bugs in static analysis rules of the bug finders, we further examine the source code of these static bug finders and localize the faults for these buggy rules listed in Tables~\ref{tab:bug_rule_implement}, \ref{tab:bug_rule_define} and \ref{tab:bug_false_positive}.
Based on our analysis of the bug localization results, we summarize the following three types of typical faults. 

\textbf{\textit{1) Inflexible design of rule implementation}}

We notice that the implementation of 60\% (15/25) detected buggy rules in PMD involves the XPath technology \cite{link_xpath_pmd}, which searches for specific expression on the Abstract Syntax Tree (AST) of the analyzed program.
This implementation is less flexible and very sensitive to noisy data.

Take the rule \textit{SimplifyConditional} from PMD as an example, we have presented the analysis in Section \ref{subsec_result_bugs_FN_imp}.P6 where we show this rule fails when AST changes by adding the unnecessary brackets.
When implementing this rule, as shown below, it would search the conditional statement (i.e., \textit{ConditionalAndExpression}) and check whether it is the EqualityExpression (i.e., \textit{rhs != null}) and InstanceOfExpression (i.e., \textit{rhs instanceof CodeLocation}) respectively.
When the two expressions are no longer in parallel after adding the brackets, the rule would fail to work.

\begin{lstlisting}[language=XML]
<![CDATA[ //Expression
[ConditionalOrExpression ....
or ConditionalAndExpression
    [EqualityExpression[@Image='!=']//NullLiteral and 
    InstanceOfExpression[PrimaryExpression[count(PrimarySuffix[@ArrayDereference='true'])=0]//Name[not(contains(@Image,'.'))]/@Image = ancestor::
    ConditionalAndExpression/EqualityExpression/PrimaryExpression/PrimaryPrefix/Name/@Image] and (count(InstanceOfExpression) + 1 = count(*))
]]]]>
\end{lstlisting}

\textbf{\textit{2) Uncovering potential influenced statements}}

The four static bug finders employ similar strategies to implement the rules.
In detail, given a specific rule, these bug finders
first categorize all the statements in the code under analysis according to their functionalities, e.g, variable definition, variable assignment, conditional judgment, etc. Then they further visit the pre-defined potential problematic statements, and analyze them to determine whether there is a match, i.e., a warning is given. 

Take the rule \textit{IntLongMath} from ErrorProne (mentioned in Section \ref{subsec_result_bugs_FN_imp}.P3 as an example, the rule first locates the problematic statements, i.e., return, initialization, assignment (i.e., \textit{matchAssignment()} as shown in the code below).
It then determines whether the result is \textit{Long} type, and filter out the operations which can not result in overflow.
However, when locating the problematic statements, it does not consider the \textit{compare statement} which can also trigger the bug.

\begin{lstlisting}
@Override
public Description matchAssignment(AssignmentTree tree, VisitorState state) {
    return check(ASTHelpers.getType(tree), tree.getExpression());
}
Description check(Type targetType, ExpressionTree init) {
    if (targetType.getKind() != TypeKind.LONG) {
      return NO_MATCH;
    }
    ...
}
\end{lstlisting}

\textbf{\textit{3) Missing considering special cases}}

Many faults are caused because of their neglecting in special cases. 
For the rule \textit{Objects should not be created only to getClass} of SonarQube, it fails in the \textit{Array} type as mentioned in Section \ref{subsec_result_bugs_FN_imp}.P1.
The implementation code shown below demonstrates that this rule is designed for \textit{java.lang.Object}, which does not include \textit{Array}.
More special cases should be included to ensure the robustness of these rules.

\begin{lstlisting}
protected List<MethodMatcher> getMethodInvocationMatchers(){
    return Collections.singletonList(MethodMatcher.create().typeDefinition(TypeCriteria.subtypeOf("java.lang.Object")).name("getClass").withoutParameter());
}
\end{lstlisting}




\subsection{Usefulness Evaluation}

To further demonstrate the usefulness of this study, for each detected bug, we create a bug report by describing the issue, the example code, and the analyzed reason, then report it to the development team through an issue report. Among the {\bug} detected bugs, {\cbug} have been fixed or confirmed by the developers, and the left are awaiting confirmation.
The fixed/confirmed bugs are marked in Table \ref{tab:bug_rule_implement} to Table \ref{tab:bug_false_positive}, and all the reported issues are listed on our website\footnote{https://github.com/wuchiuwong/Diff-Testing-01}.
The fixed and confirmed bugs further demonstrate the usefulness of this study in helping the developers improving these widely-used bug finders.

%% file: tab/bug_rule_implement.tex
\begin{table*}[ht!]
\caption{False negative bugs (about rule implementation) in the examined static bug finders}
\centering
\footnotesize
\begin{tabular}{p{1.6cm}|p{3.5cm}|p{5.5cm}|p{5.5cm}}
\hline
\textbf{Bug pattern (\# involved rules)} & \textbf{Description} & \textbf{Example} & \textbf{Involved rules (number of inconsistencies)} / \texttt{\textbf{C}} indicates fixed/confirmed bug \\ 
\hline
\multirow{5}{*}{\shortstack[l]{\textbf{\textit{P1.} Fail in}\\ \textbf{special}\\ \textbf{data type (5)}}} & \multirow{5}{*}{\shortstack[l]{Rules fail to warn the\\ suspicious code involving \\special data types}} & \multirow{5}{*}{\shortstack[l]{Rule \textit{Object should not be created only to} \\ \textit{getClass} fails with Array, while works with \\ArrayList;}} & \textit{PMD}: SingularField (327)  \\
 &  &  & \textit{SonarQube}: String function use should be optimized for single characters (8)  / \texttt{\textbf{C}}\\
 &  &  & \textit{SonarQube}: Objects should not be created only to getClass (5)  / \texttt{\textbf{C}}\\
 &  & & \textit{PMD}: RedundantFieldInitializer (3)  / \texttt{\textbf{C}}\\
 & & & \textit{SpotBugs}: ICAST\_BAD\_SHIFT\_AMOUNT (3) \\
 \hline
\multirow{5}{*}{\shortstack[l]{\textbf{\textit{P2.} Fail in}\\ \textbf{compound}\\ \textbf{expression}\\ \textbf{(5)}}} & \multirow{5}{*}{\shortstack[l]{Rules fail to warn the \\suspicious code involving \\compound expression}} & \multirow{5}{*}{\shortstack[l]{Rule \textit{QBA\_QUESTIONABLE \_BOOLEAN}\\ \textit{\_ASSIGNMENT} fails when compound\\ expression is involved, e.g., \\
	\textit{if (scanning = b == false)};}
} & \textit{ErrorProne}: ToStringReturnsNull  (28)  \\
 & & & \textit{SpotBugs}: QBA\_QUESTIONABLE\_BOOLEAN \_ ASSIGNMENT  (26)   \\
 & & & \textit{SpotBugs}: SA\_LOCAL\_SELF\_ASSIGNMENT  (26) \\
 & &  & \textit{SonarQube}: Variables should not be self-assigned (5)\\
  & & &  \textit{SonarQube}: Fields should not be initialized to default values  (3)  / \texttt{\textbf{C}}\\
\hline
\multirow{4}{*}{\shortstack[l]{\textbf{\textit{P3.} Fail in}\\\textbf{implicit}\\ \textbf{operation} \\ \textbf{(4)}}} & \multirow{4}{*}{\shortstack[l]{Rules fail to warn the \\suspicious involving the \\implicit operation of the \\defined suspicious operation}} & \multirow{4}{*}{\shortstack[l]{Rule \textit{IntLongMath (Expression of type int may} \\\textit{ overflow before assigning to a long)} fails when \\involving comparison operation (i.e.,\\ implicit assignment operation), e.g., \\
\textit{if (score $>=$ level $\times$ level $\times$ 1000);} }} & \textit{ErrorProne}: IntLongMath (135)  / \texttt{\textbf{C}}  \\
 & & & \textit{SonarQube}: Static fields should not be updated in constructors (72)  / \texttt{\textbf{C}} \\
  & & & \textit{SpotBugs}: DMI\_INVOKING\_TOSTRING\_ON\_ ARRAY (53) \\
 & & &  \textit{ErrorProne}: ArrayEquals (11)  \\
 \hline
\multirow{4}{*}{\shortstack[l]{\textbf{\textit{P4.} Fail in}\\ \textbf{multiple}\\ \textbf{calling}\\ \textbf{operations (4)}}} & \multirow{4}{*}{\shortstack[l]{Rules fail to warn the \\suspicious code involving \\multiple calling operations}} & \multirow{4}{*}{\shortstack[l]{Rule \textit{AvoidPrintStackTrace} fails when\\ involving multiple calling operations, e.g., \\
\textit{json.exception().printStackTrace();} }} & \textit{PMD}: UseCollectionIsEmpty (399)  / \texttt{\textbf{C}}\\
 & & & \textit{PMD}: AvoidPrintStackTrace (20)  / \texttt{\textbf{C}} \\
 & & & \textit{PMD}: ClassCastExceptionWithToArray (5) \\
 & & & \textit{PMD}: DontCallThreadRun (5)  / \texttt{\textbf{C}}\\
 & & & \\
\hline
\multirow{3}{*}{\shortstack[l]{\textbf{\textit{P5.} Fail in}\\ \textbf{separated}\\ \textbf{expressions}\\ \textbf{(3)}}} & \multirow{3}{*}{\shortstack[l]{Rules fail to warn the \\suspicious code involved in \\separated expressions}} & \multirow{3}{*}{\shortstack[l]{Rule \textit{UseProperClassLoader} fails when\\ involving separated expressions, \\e.g., 
\textit{Foo foo = new Foo();}  \\
\textit{ClassLoader classLoader = foo.getClassLoader();}}} & \textit{PMD}: UseProperClassLoader (68)  / \texttt{\textbf{C}}  \\
& & & \textit{ErrorProne}: ToStringReturnsNull (28)  / \texttt{\textbf{C}} \\
& & & \textit{PMD}: InstantiationToGetClass (19)  \\
& & & \\
& & & \\
\hline
\multirow{4}{*}{\shortstack[l]{\textbf{\textit{P6.} Fail in}\\ \textbf{unnecessary}\\ \textbf{brackets (3)}}} & \multirow{4}{*}{\shortstack[l]{Rules fail to warn the \\suspicious code involving \\unnecessary brackets which \\changes AST of the code }} & \multirow{4}{*}{\shortstack[l]{Rule \textit{SimplifyConditional} fails with\\ \textit{if(rhs != null \&\& (rhs instanceof} \\\textit{ CodeLocation))}, yet works when deleting the \\unnecessary brackets on 2rd condition;}}   &  \textit{PMD}: ReturnEmptyArrayRatherThanNull (770) \\
 & & & \textit{PMD}: SimplifyConditional (109) \\
 & & & \textit{SonarQube}: Switch statements should not contain non-case labels (39) / \texttt{\textbf{C}}\\
  & & & \\
 \hline
\multirow{3}{*}{\shortstack[l]{\textbf{\textit{P7.} Fail in}\\ \textbf{variables (2)}}} & \multirow{3}{*}{\shortstack[l]{Rules fail to warn the \\suspicious code involving \\variables, while works \\with constant}} & \multirow{3}{*}{\shortstack[l]{Rule \textit{FinalFieldCouldBeStatic} fails when \\assigning to an expression, e.g., \textit{private final double }\\\textit{HPI = Math.PI $*$ 0.5}, yet works when assigning to \\a constant, e.g., \textit{protected final int margin = 3};}} & \textit{PMD}:	FinalFieldCouldBeStatic (209) \\
 & & & \textit{PMD}: AvoidDecimalLiteralsInBigDecimalConstructor (20)  / \texttt{\textbf{C}}\\
  & & &  \\
 & & &  \\
 \hline
\multirow{2}{*}{\textbf{Others (3)}} & \multirow{2}{*}{\textbf{Others}} & \multirow{2}{*}{\shortstack[l]{N/A}}  & \textit{PMD}: AvoidThrowingNullPointerException (772) \\
 & & & \textit{PMD}: SimplifyBooleanExpressions (623) / \texttt{\textbf{C}}  \\
 & & & \textit{PMD}: StringToString (215)  / \texttt{\textbf{C}}\\
\hline
\end{tabular}
\label{tab:bug_rule_implement}
\end{table*}

%% file: tab/bug_rule_definition.tex
\begin{table*}[ht!]
\caption{False negative bugs (about rule definition) in the examined static bug finders.}
\centering
\footnotesize
\begin{tabular}{p{1.6cm}|p{3.2cm}|p{4.8cm}|p{6.3cm}}
\hline
\textbf{Bug pattern (\# involved rules)} & \textbf{Description} & \textbf{Example} & \textbf{Involved rules (number of inconsistencies)} / \texttt{\textbf{C}} indicates fixed/confirmed bug \\ 
\hline
\multirow{4}{*}{\shortstack[l]{\textbf{\textit{P8.} Miss}\\ \textbf{ comparable} \\ \textbf{method (4)}}} & \multirow{4}{*}{\shortstack[l]{Rules miss comparable \\method of the defined \\suspicious method}} & \multirow{4}{*}{\shortstack[l]{Rule \textit{UseLocaleWithCaseConversions} fails with\\ \textit{String.format()}, while works with\\  \textit{String.toLowerCase()/toUpperCase();}}} & \textit{PMD}: UseLocaleWithCaseConversions (464)  \\
 &  & & \textit{ErrorProne}: BoxedPrimitiveEquality (14)  \\
 & & & \textit{SonarQube}: Java.lang.Error should not be extended (7)  / \texttt{\textbf{C}} \\
  &  & & \textit{SonarQube}: Execution of the Garbage Collector should be triggered only by the JVM (7) / \texttt{\textbf{C}} \\
 \hline
 \multirow{3}{*}{\shortstack[l]{\textbf{\textit{P9.} Miss}\\ \textbf{comparable}\\ \textbf{data type or}\\ \textbf{operation (3)}}} & \multirow{3}{*}{\shortstack[l]{Rules miss the comparable \\data type or operation \\of the defined suspicious ones}} & \multirow{3}{*}{\shortstack[l]{Rule \textit{ICAST\_INTEGER\_MULTIPLY\_CAST} \\ \textit{\_TO\_LONG} fails with \textit{shift} operation,\\ while works with \textit{multiply} operation; }} & \textit{SonarQube}: Redundant modifiers should not be used (1271)  / \texttt{\textbf{C}}\\
& & & \textit{PMD}: AvoidArrayLoops (362) / \texttt{\textbf{C}} \\
 &  & & \textit{SpotBugs}: ICAST\_INTEGER\_MULTIPLY\_CAST\_TO\_LONG (39) \\
&  & & \\
 \hline
\multirow{2}{*}{\shortstack[l]{\textbf{\textit{P10.} Miss}\\ \textbf{subclass or} \\ \textbf{superclass (2)}}} & \multirow{2}{*}{\shortstack[l]{Rules miss the subclass or \\superclass of the defined \\suspicious class}} & \multirow{2}{*}{\shortstack[l]{Rule \textit{AvoidCatchingThrowable} fails in the \\subclass of \textit{Throwable}, e.g., catch (Error e);}} & \textit{PMD}: ReturnEmptyArrayRatherThanNull (677) \\
 &  & & \textit{PMD}: AvoidCatchingThrowable (41)   \\
  &  & &   \\
\hline
\end{tabular}
\label{tab:bug_rule_define}
\end{table*}

%% file: tab/bug_false_positive.tex
\begin{table*}[ht!]
\caption{False positive bugs in the examined static bug finders}
\centering
\footnotesize
\begin{tabular}{p{1.85cm}|p{3.5cm}|p{5.8cm}|p{5.2cm}}
\hline
\textbf{Bug pattern (\# involved rules)} & \textbf{Description} & \textbf{Example} & \textbf{Involved rules (number of inconsistencies)} / \texttt{\textbf{C}} indicates fixed/confirmed bug \\ 
\hline
\multirow{3}{*}{\shortstack[l]{\textbf{\textit{P11.} Poor}\\ \textbf{handling of}\\ \textbf{method with} \\\textbf{same name (3)}}} & \multirow{3}{*}{\shortstack[l]{
Rules wrongly warn the clean method \\which has the same name yet \\different method signatures with \\the defined suspicious method}} & \multirow{3}{*}{\shortstack[l]{ \textit{UseNotifyAllInsteadOfNotify} should warn\\ \textit{Object.notify()}, but wrongly warns \textit{notify(para)} \\method in other class, \\e.g., getEventThread().notify(input);}} & \textit{PMD}: SuspiciousEqualsMethodName (24)\\
 &  & & \textit{PMD}: UseNotifyAllInsteadOfNotify (12) / \texttt{\textbf{C}} \\
& & & \textit{SonarQube}: Object.finalize() should remain protected when overriding (6) / \texttt{\textbf{C}} \\
& & & \\
 \hline
\multirow{3}{*}{\shortstack[l]{\textbf{\textit{P12.} Setting}\\ \textbf{over-sized}\\ \textbf{scope (3)}}} & \multirow{3}{*}{\shortstack[l]{Rules wrongly warn the clean\\ code which is beyond the scope of \\the defined suspicious case}} & \multirow{3}{*}{\shortstack[l]{\textit{AvoidThrowingNullPointerException} wrongly warns the \\expression with \textit{NullPointerException} yet without \textit{Throwing}, \\e.g., Exception e = new NullPointerException(``msg'');}} & 
\textit{PMD}: AvoidThrowingNullPointerException (18) / \texttt{\textbf{C}} \\
 &  & & \textit{SpotBugs}: SF\_SWITCH\_NO\_DEFAULT (11)  \\
& & & \textit{PMD}: AvoidCallingFinalize (6) / \texttt{\textbf{C}} \\
& & & \\
 \hline
\multirow{2}{*}{\shortstack[l]{\textbf{\textit{P13.} Neglecting}\\ \textbf{corner}\\ \textbf{case (2)}}} & \multirow{2}{*}{\shortstack[l]{Rules wrongly warn the clean\\ code which is the corner case\\ of the defined suspicious case}} & \multirow{2}{*}{\shortstack[l]{\textit{MissingBreakInSwitch} wrongly warns the\\ \textit{switch} expression without \textit{break} in the\\ last \textit{case} statement;}} & 
\textit{PMD}: MissingBreakInSwitch (841) / \texttt{\textbf{C}} \\
 &  & & \textit{PMD}: AvoidReassigningLoopVariables (188) \\
 & & & \\
\hline
\end{tabular}
\label{tab:bug_false_positive}
\end{table*}

%% file: sec/discussion.tex
\section{Discussions and Threats to Validity}
\label{sec_discussion}

\subsection{Discussions}
\label{subsec_discussion}

\textbf{Checklist for static analysis rule design and implementation.}
Table \ref{tab:bug_rule_implement}, \ref{tab:bug_rule_define}, and \ref{tab:bug_false_positive} summarizes the bug patterns in the implementation and design of static analysis rules.  
We also notice that most of these bug patterns involve rules target at different types of warnings and from different bug finders, which implies the generalizability of these bug patterns. 

The bug patterns are actually the special cases which is ignored by the static analysis rules and cause them fail to warn the corresponding suspicious code or wrongly warn the clean code. 
We believe these bug patterns can serve as the checklist when one designs or implements new rules and/or in other static bug finders.
Take bug pattern \textit{P2. Fail in compound expression} as an example, we find some static analysis rules fail to warn the suspicious code involving compound expression.
Equipped with such a bug pattern, one should pay careful attention to this special case when designing/implementing a new rule, and include related test cases to cover this special case, both of which can help improve the quality of newly implemented static analysis rules.

\textbf{Customizing static bug finders to avoid execution of duplicate rules (i.e., paired rules).}
Many software organizations tend to configure multiple bug finders for detecting bugs with an assumption that different bug finders emphasize on detecting different types of bugs in the source code \cite{lu2018evaluating,Sadowski2018lessons}.
For example, Github projects as Springfox\footnote{https://github.com/springfox/springfox} and Roboguice\footnote{https://github.com/roboguice/roboguice} employ both SpotBugs and PMD for code inspection \cite{Zampetti2017how}. 
Another example is SonarQube which incorporates static analysis rules from other bug finders, i.e., SpotBugs, PMD, cobertura, and CheckStyle (note that, we exclude these external rules in our experiment).

Our rule mapping results reveal a non-negligible portion of duplicate rules among the examined static bug finders, e.g., at least 24\% (74/304) rules from PMD are the duplicates of the rules in SonarQube. 
However, most of the bug finders are used in default configuration \cite{Vassallo2020how}, which results in the duplicate rules repeatedly running and brings in heavy overhead.   
A more feasible alternative would be customizing these bug finders to make the duplicate rules only execute once, and the mapped rule pairs retrieved in this study further provide the feasibility for the customization.
The differential testing results of this study also provide the detailed guidelines to do so, e.g., choose the correct rule if one of them is buggy.


 




\textbf{Multiple bug patterns of one rule or two paired rules.}
We notice that one static analysis rule (or two rules in a pair) can involve multiple bug patterns, i.e., having different types of bugs. 
For example, the rule \textit{AvoidThrowingNullPointerException} of PMD is listed in both false negative bugs about rule implementation and false positive bugs. 
In the former case, the rule can not warn the suspicious code which throws the exception in method definition (see Table \ref{tab:bug_rule_implement}), while in the latter case, the rule wrongly warns the normal code with \textit{NullPointerException} yet without \textit{throw} operation (see details in Section \ref{subsec_result_bugs_FP}.P12).

The second example is for the paired rules, i.e.,  \textit{IntLongMath} of ErrorProne and \textit{ICAST\_INTEGER\_MULTIPLY\_CAST\_TO\_LONG} of SpotBugs.
The former fails with the \textit{assignment} operation (see details in Section \ref{subsec_result_bugs_FN_imp}.P3), and is listed in false negative bugs about pattern implementation.
The latter only considers the \textit{multiply} operation in the rule definition, while misses other related operation as \textit{shift} (see details in Section \ref{subsec_result_bugs_FN_def}.P9).
These diversified cases increase the difficulty of testing these static analysis rules and further indicates the value of this study. 

\subsection{Threats to Validity}
\label{subsec_method_threats}


The first threat of this study is the selection of static bug finders, which may or may not be representative for a larger population.
We experiment with four popular open source static bug finders, which are widely used in previous researches and industrial practice \cite{bessey2010few,Ayewah2008using,wang2018is,Zampetti2017how}, which we believe to be representative for the current state-of-the-art. 
Despite of this, there are other commonly-used static bug finders, e.g., Infer, CheckStyle, Coverity.
The reason why do not utilize Infer or CheckStype is because they either have few static analysis rules (e.g., 25 rules in Infer) 
or focus more on the coding standards (i.e., CheckStyle), both of which limit us in finding plenty of mapping rules and detecting more inconsistencies. 
Besides, Coverity is a closed sourced bug finder with which we could not conduct the bug localization.

Another threat to validity is our methodology for mapping static analysis rules which could, in principle, miss some paired rules. 
To overcome the challenges that tens of thousands of candidate pairs needed to be examined, we design a heuristic-based rule mapping method. 
This could miss some true paired rules, yet make this mapping task can be done with reasonable human effort. 
We employ several empirical parameters in filtering the candidate rule pairs which might also influence the recall of paired rules.
Nevertheless, based on the retrieved mapping rules, we have detected {\bug} bugs in the tools.
Besides, we present the hitting rate and recall of rule mapping in Section \ref{subsec_results_mapping} to show the reliability of the method.


In addition, this study does not include the one-to-many mapping rule combinations, since its potential candidate number is far too large, and there would be correspondingly more confusing cases. 
Future work will adopt other mechanisms as multi-input learning, to help automatically retrieve such kinds of rule combinations, so as to conduct differential testing and detect bugs on more rules.


 






%% file: sec/related.tex
\section{Related Work}
\label{sec_related}


The use of static bug finders for software defect detection is a common practice for developers during software development and has been studied by many researchers \cite{flanagan2002extended,Nagappan2005static,zheng2006on,Rahman2014comparing,johnson2013why}.
There were studies investigating the adoption of static bug finders within continuous integration pipelines \cite{Zampetti2017how,Panichella2015would,Vassallo2018continuous}. 
Other studies focused on how these warnings are actually acted, and fixed \cite{Marcilio2019are,Imtiaz2019how,Thung2015what}.
Several researchers proposed to utilize prioritization strategies to make it easier for the developers to spot the more actionable warnings \cite{heckman2008establishing,kim2007which,liang2010automatic,heckman2011systematic,wang2018is}.
There were several researches employing feedback-based rule design for mitigating false positives \cite{Sadowski2015Tricorder,Nam2019bug}.
Another line of researches focused on designing project-specific rules that mined from specific projects to improve the detection accuracy \cite{li2005PRminer,Livshits2005dynamine,jin2012understanding,hanam2016discovering,Chen2017characterizing,bian2018NARminer}.
All above mentioned researches focused on improving current static bug finders by adjusting the warning results or designing new rules, while this study investigate the correctness of bug finders which can improve the detection accuracy fundamentally.

Differential testing is originally introduced by McKeeman which attempts to detect bugs by checking inconsistent behaviors across different comparable software or different software versions \cite{mckeeman1998differential}. 
Randomized differential testing is a widely-used black-box differential testing technique in which the inputs are randomly generated \cite{Yang2011findi,tian2019differential,Lidbury2015manycore,le2014compiler,sun2016finding,pham2019cradle,guo2018dlfuzz,yang2019hunting,chen2019deep}.
Inconsistency detection has been used in other domains such as cross-platform \cite{Fazzini2017automated}, web browsers \cite{Choudhary2011detecting}, document readers \cite{Kuchta2018correctness}, and program variables  \cite{Sayali2018phys}. 
Different from the above mentioned application scenarios, we apply differential testing to detect bugs in widely-used static bug finders which can improve the performance of bug detection.

%% file: sec/conclusion.tex
\section{Conclusion}
\label{sec_conclusion}

Static bug finders are widely used by professional software developers, and regularly integrated in contemporary open source projects and commercial software organizations. 
They have been shown to be helpful in detecting software defects faster and cheaper than human inspection or software testing. 
To further improve the reliability of static bug finders, this paper proposes a differential testing approach to detect bugs in the static analysis rules of four widely-used static bug finders. 
Our study finds {\bug} bugs about the implementation or design of static analysis rules, among which {\cbug} are fixed/confirmed and the left are awaiting confirmation. 
We also summarize {\pattern} bug patterns in the static analysis rules based on their context and root causes, which can serve as the checklist for designing and implementing other rules.

The detected bugs provide a strong evidence that the commonly-used static bug finders are not as reliable as they might have been envisaged. 
Follow-up studies are encouraged to conduct testing on other popular static bug finders, or cover more rules of the four examined bug finders in this paper. 
This can be done following our proposed differential testing approach, and the heuristic-based static analysis rule mapping method.